\documentclass[12pt]{article} 
\usepackage[sectionbib]{natbib}
\usepackage{array,epsfig,fancyheadings,rotating}
\usepackage[]{hyperref}  
\usepackage{sectsty, secdot}
\sectionfont{\fontsize{12}{14pt plus.8pt minus .6pt}\selectfont}
\renewcommand{\theequation}{\thesection\arabic{equation}}
\subsectionfont{\fontsize{12}{14pt plus.8pt minus .6pt}\selectfont}

\usepackage{amsmath}
\usepackage{amssymb}
\usepackage{amsfonts}
\usepackage{multirow}
\usepackage{amsthm}
\usepackage{cleveref}
\usepackage[font=small, skip=-7pt]{caption}

\newtheorem{innercustomgeneric}{\customgenericname}
\providecommand{\customgenericname}{}
\newcommand{\newcustomtheorem}[2]{%
	\newenvironment{#1}[1]
	{%
		\renewcommand\customgenericname{#2}%
		\renewcommand\theinnercustomgeneric{##1}%
		\innercustomgeneric
	}
	{\endinnercustomgeneric}
}

\newcustomtheorem{customcondition}{Condition}

\newtheorem{theorem}{Theorem}
\newtheorem{lemma}{Lemma}

\newtheorem{proposition}{Proposition}
\newtheorem{condition}{Condition}
\theoremstyle{definition}

\usepackage{mathrsfs}
\usepackage{natbib, graphicx, color, setspace, amssymb, amsmath, amsthm, amstext, booktabs, fancyhdr, multirow, float, bm, lineno, xr, subfigure,enumerate,url}
\usepackage{times}
\allowdisplaybreaks 
\usepackage{pdflscape}
\usepackage{placeins} 
\usepackage{authblk} 
\usepackage{longtable}
\usepackage[ruled,vlined,onelanguage]{algorithm2e}
\usepackage{comment}

\usepackage{xcite}
\makeatletter
\newcommand*{\addFileDependency}[1]{
  \typeout{(#1)}
  \@addtofilelist{#1}
  \IfFileExists{#1}{}{\typeout{No file #1.}}
}
\makeatother

\newcommand*{\myexternaldocument}[1]{%
    \externaldocument{#1}%
    \addFileDependency{#1.tex}%
    \addFileDependency{#1.aux}%
}

\myexternaldocument{Supplementary}

\newcommand*\patchAmsMathEnvironmentForLineno[1]{%
	\expandafter\let\csname old#1\expandafter\endcsname\csname #1\endcsname
	\expandafter\let\csname oldend#1\expandafter\endcsname\csname end#1\endcsname
	\renewenvironment{#1}%
	{\linenomath\csname old#1\endcsname}%
	{\csname oldend#1\endcsname\endlinenomath}}%
\newcommand*\patchBothAmsMathEnvironmentsForLineno[1]{%
	\patchAmsMathEnvironmentForLineno{#1}%
	\patchAmsMathEnvironmentForLineno{#1*}}%
\AtBeginDocument{%
	\patchBothAmsMathEnvironmentsForLineno{equation}%
	\patchBothAmsMathEnvironmentsForLineno{align}%
	\patchBothAmsMathEnvironmentsForLineno{flalign}%
	\patchBothAmsMathEnvironmentsForLineno{alignat}%
	\patchBothAmsMathEnvironmentsForLineno{gather}%
	\patchBothAmsMathEnvironmentsForLineno{multline}%
}
\usepackage{etoolbox}
\usepackage{tikz}
\usepackage{array}

\newcommand{\bmalpha}{\bm{\alpha}}

\newcommand{\bmtheta}{\bm{\theta}}

\newcommand{\bmW}{\bm{W}}

\newcommand{\bmalphao}{\bmalpha^{(0)}}

\newcommand{\alphao}{\alpha^{(0)}}

\newcommand{\bmvartheta}{\bm{\vartheta}}

\DeclareMathAlphabet{\mathbcal}{OMS}{cmsy}{b}{n}

\def\beq{\begin{equation}}
	\def\eeq{\end{equation}}
\def\beqr{\begin{eqnarray}}
	\def\eeqr{\end{eqnarray}}
\def\beqrs{\begin{eqnarray*}}
	\def\eeqrs{\end{eqnarray*}}
\def\bet{\begin{theorem}}
	\def\eet{\end{theorem}}
\def\bel{\begin{lemma}}
	\def\eel{\end{lemma}}
\def\bep{\begin{proposition}}
	\def\eep{\end{proposition}}
\def\bg{\begin{figure}[tbph]\begin{center}}
		\def\eg{\end{center}\end{figure}}

\def\bc{\begin{center}}
	\def\ec{\end{center}}

\def\bmmX{\bm{\mathcal{X}}}

\def\be{\begin{equation}}
	\def\ee{\end{equation}}
\def\ben{\begin{equation*}}
	\def\een{\end{equation*}}
\def\bea{\begin{eqnarray}}
	\def\eea{\end{eqnarray}}

\def\bda{\begin{eqnarray*}}
	\def\eda{\end{eqnarray*}}

\def\lsk{\left(}
\def\rsk{\right)}
\def\lbk{\left \{ }
\def\rbk{\right \} }

\def\lak{\left | }
\def\rak{\right | }
\def\laak{\left \| }
\def\raak{\right \| }

\newcommand{\zy}[1]{{\leavevmode\color{black}#1}}

\begin{document}
	
	
	\renewcommand{\baselinestretch}{2}
	
	\markright{ \hbox{\footnotesize\rm 
		}\hfill\\[-13pt]
		\hbox{\footnotesize\rm
		}\hfill }
	
	\markboth{\hfill{\footnotesize\rm Zhi Yang Tho AND Francis K.C. Hui AND Tao Zou} \hfill}
	{\hfill {\footnotesize\rm Ising Similarity Regression Model} \hfill}
	
	\renewcommand{\thefootnote}{}
	$\ $\par
	
	\renewcommand{\thefootnote}{*} 
	\fontsize{12}{14pt plus.8pt minus .6pt}\selectfont \vspace{0.8pc}
	\centerline{\large\bf An Ising Similarity Regression Model}
	\vspace{2pt} 
	\centerline{\large\bf for Modeling Multivariate Binary Data}
	\vspace{.4cm} 
	\centerline{Zhi Yang Tho\footnote{Corresponding author.}, Francis K. C. Hui, and Tao Zou\thanks{Corresponding author.}} 
	\vspace{.4cm} 
	\centerline{\it The Australian National University}
	\vspace{.55cm} \fontsize{9}{11.5pt plus.8pt minus.6pt}\selectfont
	
	
	\begin{quotation}
		\noindent {\it Abstract:}
		Understanding the dependence structure between response variables is an important component in the analysis of correlated multivariate data. This article focuses on modeling dependence structures in multivariate binary data, motivated by a study aiming to understand how patterns in different U.S. senators' votes are determined by similarities (or lack thereof) in their attributes, e.g., political parties and social network profiles. To address such a research question, we propose a new Ising similarity regression model which regresses pairwise interaction coefficients in the Ising model against a set of similarity measures available/constructed from covariates.  Model selection approaches are further developed through regularizing the pseudo-likelihood function with an adaptive lasso penalty to enable the selection of relevant similarity measures. We establish estimation and selection consistency of the proposed estimator under a general setting where the number of similarity measures and responses tend to infinity. Simulation study demonstrates the strong finite sample performance of the proposed estimator, \zy{particularly compared with several existing Ising model estimators in estimating the matrix of pairwise interaction coefficients.}
        Applying the Ising similarity regression model to a dataset of roll call voting records of 100 U.S. senators, we are able to quantify how similarities in senators' parties, businessman occupations and social network profiles drive their voting associations.

		\vspace{9pt}
		\par
	\end{quotation}\par

	\def\thefigure{\arabic{figure}}
	\def\thetable{\arabic{table}}
	
	\renewcommand{\theequation}{\thesection.\arabic{equation}}

	\fontsize{12}{14pt plus.8pt minus .6pt}\selectfont
	
	\renewcommand{\thefootnote}{\arabic{footnote}}

\section{Introduction} \label{sec:intro}

The study of the dependence structure in correlated multivariate data  has drawn much attention in recent years, due to the increasing accessibility of multi-response data across many disciplines such as finance \citep{tsay2013multivariate}, economics \citep{mcelroyANDtrimbur2023} and ecology \citep*{huiETAL2023gee}. For a multivariate response vector $\bm{y} = (y_1,\cdots,y_p)^\top$, the dependencies across any set of two components $y_{j}$ and $y_{j'}$ for $j,j'=1,\cdots,p$ can be modeled by specifying a joint probabilistic distribution for $\bm{y}$. Two prominent examples are Gaussian graphical models  \citep{whittaker1990} and Ising models \citep{ising1925}, which specify a probability density function (pdf)  and  a probability mass function (pmf) for a continuous and binary  response vector $\bm{y}$, respectively, where the pdf and pmf of these two models both contain a component of the form $\sum_{1\leq j<j'\leq p}\exp(\theta_{jj'}y_{j}y_{j'})$ with $\theta_{jj'}$ denoting a coefficient for the interaction term between $y_{j}$ and $ y_{j'}$.  Accordingly, the set of \{$\theta_{jj'}: j,j'=1,\cdots,p$\} are referred to as  pairwise interaction coefficients in these joint models. Moreover, in both models each $\theta_{jj'}$ is directly tied to the conditional dependence relationship between $y_{j}$ and $y_{j'}$ given the remaining elements in the response vector $\bm{y}$ \citep[Hammersley-Clifford equivalence,][]{hammersleyANDclifford1971}: if $\theta_{jj'}$ is close to/far away from zero, the conditional dependence between $y_{j}$ and $y_{j'}$ is weak/strong, with the limiting case of $\theta_{jj'} = 0$ implying conditional independence. This interpretation of the pairwise interaction coefficient $\theta_{jj'}$ has contributed to the popularity of Gaussian graphical models and Ising models for correlated multivariate data analysis; see for instance  \citet{meinhausenANDbuhlmann2006}, \citet{yuanANDlin2007} and \citet*{friedmanETAL2008} for the Gaussian graphical models, and  \citet*{majewskiETAL2001}, \citet{guoETAL2015} and \citet{bhattacharyaANDmukherjee2017} for the Ising models, among many others. 

In this article, we focus on modeling the dependence relationships for a multivariate binary response vector $\bm{y}$ through, or equivalently, modeling the  pairwise  interaction coefficients $\theta_{jj'}$ in, the Ising model. As demonstrated above, the conditional dependence of multivariate binary responses in the Ising model is captured by the pairwise interaction coefficients $\theta_{jj'}$, which can be collected into a symmetric interaction matrix $\bm{\Theta} = (\theta_{jj'})_{p \times p}$ where $\theta_{j'j} = \theta_{jj'}$. A variety of approaches have been proposed to estimate and regularize $\bm{\Theta} $. For example, \citet*{banerjeeETAL2008} proposed a block-coordinate descent algorithm to solve an approximate sparse maximum likelihood problem  for the estimation of  $\bm{\Theta}$. \citet*{ravikumarETAL2010} considered a neighborhood  estimation method based on fitting a separate regularized logistic regression with the lasso penalty to each binary response, while \citet{hoflingANDtibshirani2009} and \citet{guoETAL2010} used a pseudo-likelihood function coupled with a lasso-type penalty to simultaneously estimate  and regularize all pairwise interaction  coefficients in $\bm{\Theta}$. Further penalized likelihood approaches for sparse $\bm{\Theta}$ estimation have been developed by \citet*{leeETAL2006} and \citet*{xueETAL2012}, among others.

In contrast to estimating and regularizing the elements of the interaction matrix $\bm{\Theta}$, this article proposes a novel Ising similarity regression model which regresses the pairwise interaction coefficients $\theta_{jj'}$ against a set of pairwise similarity measures  $w_{jj'}^{(k)}$; that is,
\begin{equation}\label{eq:model1}
\theta_{jj'}=\sum_{k=1}^K\alpha_k w_{jj'}^{(k)},\textrm{ for }j\neq j',
\end{equation}
where $\alpha_k$ denotes the regression coefficient associated with the $k$-th similarity measure  $w_{jj'}^{(k)}$  for $k=1,\cdots,K$. 
It is worth noting $w_{jj'}^{(k)}$ measures the similarity between $j$ and $j'$, which can be either observed directly as part of the data collection process, or induced from available  auxiliary information. As a motivating example, we consider U.S. Senate roll call voting data where the binary response $y_j$ represents the $j$-th senator's voting record (Yea or Nay) to a particular piece of legislation or bill, and $w_{jj'}^{(k)}$ are constructed based on additional attributes of the $j$-th and $j'$-th senators, such as political parties and occupations. In such an example, the regression coefficients $\alpha_k$ in model \eqref{eq:model1} offer a clear, explicit quantification of how the $k$-th similarity measure between the $j$-th and $j'$-th senators affects their conditional dependence when it comes to voting patterns. 

At this point, it is important to acknowledge the work of \citet{chengETAL2014}, who proposed an Ising regression model to regress the pairwise interaction coefficients $\theta_{jj'}$  on a covariate vector $\bm{x}$. Such a model with $\theta_{jj'}=\theta_{jj'}(\bm{x})$ does not provide the same interpretation for our motivating example however, as their interaction coefficients $\theta_{jj'}$ for different pairs $(j,j')$ only depend on the same covariate vector $\bm{x}$ instead of the pairwise similarity measures $w_{jj'}^{(k)}$ considered in model \eqref{eq:model1}. \zy{\label{page:suggested_literature}Also, in a parallel line of research, several studies have considered using covariates to model the precision matrix encoding the dependence structure in Gaussian graphical models. \citet{liuETAL2010} partitioned the covariate space based on classification and regression trees into multiple subspaces that can have different precision matrices, while \citet{leeANDxue2018} proposed a nonparametric mixture of Gaussian graphical models with mixture probabilities and precision matrices that are both covariate-dependent. More recently, \citet{wangETAL2022} and \citet*{niETAL2022} proposed Bayesian approaches to model the elements of the precision matrix as linear functions of covariates. As these studies focus on precision matrices of Gaussian graphical models, they are not directly applicable to the multivariate binary response setting which we focus on in this article.}

\label{page:additional_application_scenarios}
Apart from the aforementioned example in political science, the proposed Ising similarity regression model has a multitude of applications across finance and ecology. For example, in  the study of capital markets, $y_j$ can be an indicator denoting whether the $j$-th firm has distributed dividends to its shareholders, and $w_{jj'}^{(k)}$ can be the similarity between the $j$-th and $j'$-th firms' financial fundamentals such as market value, cash flow, and leverage. Also, in ecology, $y_j$ can be a binary variable indicating the presence or absence of the $j$-th species, and $w_{jj'}^{(k)}$ can be the similarity between the $j$-th and $j'$-th species' trait values; \zy{\label{page:beetle_discussion}see an application of the proposed model to such an ecology dataset in Section \ref{sec:beetle_application} of the supplementary material, to illustrate the wide applicability of the proposed method}. More broadly, the use of similarity measures in regression is also motivated by recent developments in covariance regression modeling \citep{tao2017, tao2020,tao2021}, where the covariance between continuous responses is modeled as a linear combination of similarity measures.  \zy{\label{page:previous_link_to_literature_start}Indeed, the proposed model can be written in a matrix regression form, and this connects to the wider literature linking (functions of) matrix parameters to a linear combination of matrices; see for example \citet{anderson1973}, \citet{pourahmadi1999} and \citet{bonat2016}. To the best of our knowledge though, the proposed Ising similarity regression model is the first to establish such an idea for the Ising model specifically, by explicitly linking similarity matrices $\bm{W}_k = (w_{jj'}^{(k)})_{p \times p}$ to the interaction matrix $\bm{\Theta}$ that captures the conditional dependence of binary responses.\label{page:previous_link_to_literature_end}}
 
To estimate the proposed model, we study a regularized pseudo-likelihood approach which augments the pseudo-likelihood function of model \eqref{eq:model1} with an adaptive lasso penalty \citep{zou2006}. Doing so induces sparse estimation of the regression coefficients $\{\alpha_k: k = 1,\cdots,K\}$, which allows us to recover similarity measures that are truly relevant in explaining the conditional dependence relationships between the binary responses. \zy{\label{page:similarity_selection_intro}It is important here to highlight that this article differs from the aforementioned studies on Ising model estimation, as our main focus is to induce sparsity on the regression coefficients $\alpha_k$ as opposed to the similarity measures $w_{jj'}^{(k)}$ and the resulting interaction matrix $\bm{\Theta}$. Put another way, we aim to identify important drivers of the conditional dependence relationships i.e., similarity selection rather than edge selection, by treating the similarity measures as given covariates.}
	
Under a setting where the number of regression coefficients $K$ and responses $p$ tend to infinity, we establish estimation and  selection consistency for the regularized pseudo-likelihood estimator. To select the tuning parameter in the adaptive lasso penalty, we employ a cross-validation approach which preserves the dependencies between the elements of $\bm{y}$. Simulation results support the theoretical findings of the proposed estimator, demonstrating its strong finite sample estimation and model selection performance. \zy{\label{page:alternative_estimator_discussion_intro}Specifically, the proposed estimator not only outperforms other estimators such as the unpenalized estimator and the lasso-penalized estimator in estimating the parameters of the Ising similarity regression model, but also performs much better than the traditional Ising model estimators that ignore the additional information from similarity measures in estimating the Ising model interaction matrix $\bm{\Theta}$. Additionally, we carry out simulation studies to compare the similarity selection performance of our cross-validation approach to the use of AIC \citep{akaike1998information} and BIC \citep{schwarz1978} criteria for choosing the tuning parameter, with results showing that BIC has a comparable performance to cross-validation approach while AIC tends to suffer from overfitting.} We apply the Ising similarity regression model to roll call voting records of 100 U.S. senators from the 117-th Congress, with results demonstrating how similarities of senators' attributes and social network activities drive the association between their voting patterns. In particular, aside from the expected findings such as senators from the same state or party being more likely to vote similarly, we find that senators who are businessmen or share certain follower-followee relationships on Twitter tend to exhibit more similar voting patterns. 
	 
The rest of this article is organized as follows. Section \ref{sec:model} introduces the Ising similarity regression model along with the proposed regularized pseudo-likelihood estimator. Section \ref{sec:theory} discusses the theoretical properties of the regularized estimator. Section \ref{sec:simulation} presents simulation studies, while an application to U.S. roll call voting dataset is provided in Section \ref{sec:real_data}. Section \ref{sec:conclusion} offers some concluding remarks. All theoretical proofs of the theorems developed in this article\zy{, along with detailed empirical comparisons to other estimation approaches, as well as an additional application to Scotland Carabidae ground beetle dataset,}  are presented in the supplementary material.

 \section{An Ising Similarity Regression Model} \label{sec:model}
	
\subsection{Model Set-Up} \label{subsec:model_setup}
Let  $\bm{u} = (u_{1},\cdots,u_{p})^\top $ be any vector in the space $ \{0,1\}^p$. The Ising model \citep{ising1925}  specifies  the following pmf for the $p$-dimensional binary response vector $\bm{y}$,
\begin{align}
f(\bm{u}; \bmtheta) = P(\bm{y} = \bm{u}; \bm{\theta}) = \frac{1}{Z(\bm{\theta})} \exp \lsk \sum_{j=1}^{p} \theta_{jj} u_{j} + \sum_{1\leq j < j' \leq p} \theta_{jj'} u_{j} u_{j'} \rsk,\label{eq:ising_regression_model0}
\end{align}
where $\bmtheta = (\theta_{11},\cdots,\theta_{1p},\cdots,\theta_{p-1,p-1},  \theta_{p-1,p}, \theta_{pp})^\top$ is a $p(p+1)/2$-dimensional parameter vector, and the partition function $Z(\bm{\theta}) = \sum_{\bm{u} \in \{0,1\}^p} \exp ( \sum_{j=1}^{p} \theta_{jj} u_{j} + \sum_{1\leq j < j' \leq p} \theta_{jj'} u_{j} u_{j'} )$ is an intractable normalization constant in the pmf. The main effect parameters in \eqref{eq:ising_regression_model0} are given by $\theta_{jj}$ for $j=1,\cdots,p$, while the pairwise interaction coefficients in the pmf are given by $\theta_{jj'}$ for $j,j' = 1,\cdots,p$ and $\theta_{j'j}=\theta_{jj'}$. 
As discussed in Section \ref{sec:intro},  the symmetric interaction matrix $\bm{\Theta} = (\theta_{jj'})_{p \times p}$ is useful for studying the conditional dependence structure among binary responses.
	
\zy{\label{page:shortened_construction_W}Suppose now that, in addition to observing the binary  response vector  $\bm{y}$, we also record a set of $p \times p$ symmetric similarity matrices $\bmW_k = (w_{jj'}^{(k)})_{ p\times p}$ for $k = 1,\cdots,K$, which may be available directly as part of the data collection process or constructed from auxiliary information variables $z_{j1},\cdots,z_{jK}$ associated to the $j$-th  response for $j=1,\cdots,p$. In the latter, each element $w_{jj'}^{(k)}$  of the similarity matrix $\bmW_k$ measures the similarity between $z_{jk}$ and $z_{j'k}$ for $j\neq j'$. For instance, if $z_{jk}$ is quantitative, then we can set  $w_{jj'}^{(k)} = \exp (-|z_{jk} - z_{j'k}|^2)$, whereas if $z_{jk}$ is qualitative then we set $w_{jj'}^{(k)} = 1$ if $z_{jk}$ and $z_{j'k}$ have the same categorical level, and $w_{jj'}^{(k)} = 0$ otherwise \citep[see also][]{johnsonANDwichern1992}.}  For completeness and reasons of parameter identifiability, the diagonals $w_{jj}^{(k)}$ are set to be zeros for all $j=1,\cdots,p$ and $k = 1,\cdots,K$. 

Given a set of similarity matrices $\bmW_k$ for $k = 1,\cdots,K$, the Ising similarity regression model as introduced in equation \eqref{eq:model1} can be equivalently formulated as modeling the interaction matrix $\bm{\Theta}$ via the form
\begin{align}
\bm{\Theta} = \sum_{j=1}^{p} \theta_{jj} \bm{\Delta}_{jj} + \sum_{k=1}^{K} \alpha_k \bmW_k,
\label{eq:matrix_regression}
\end{align}
where $\bm{\Delta}_{jj}$ is a $p \times p$ matrix with the $(j,j)$-th element being one and other elements being zeros  for $j = 1,\cdots,p$. This model re-parameterizes the vector $\bmtheta$ in equation \eqref{eq:ising_regression_model0} by a new parameter vector $\bmvartheta = (\theta_{11},\cdots,\theta_{pp},\bmalpha^\top)^\top$ with $\bmalpha = (\alpha_1,\cdots,\alpha_K)^\top$, such that the re-parameterized pmf for $\bm{y}$ can be written as
	\begin{align}
		f(\bm{u};\bmvartheta) = \frac{1}{Z(\bmvartheta)} \exp \lbk \sum_{j=1}^{p} \theta_{jj} u_j + \sum_{1 \leq j < j' \leq p} \lsk \sum_{k=1}^{K} \alpha_k w_{jj'}^{(k)} \rsk u_j u_{j'} \rbk,
		\label{eq:ising_regression_model}
	\end{align}
where  $Z(\bmvartheta) = \sum_{\bm{u} \in \{0,1\}^p} \exp \{ \sum_{j=1}^{p} \theta_{jj} u_{j} + \sum_{1\leq j < j' \leq p} (\sum_{k=1}^{K} \alpha_k w_{jj'}^{(k)}) u_{j} u_{j'} \} $.  
	
In equation \eqref{eq:matrix_regression}, the vector of regression coefficients $\bmalpha$ describes how similarities directly affect the pairwise interaction coefficients (conditional dependence relationships) of the binary responses. Note also by utilizing similarity matrices to model the interaction matrix $\bm{\Theta}$, model \eqref{eq:ising_regression_model} involves a substantially smaller number of parameters ($p+K$ parameters) compared to the standard Ising model in \eqref{eq:ising_regression_model0}, which has $(p+1)p/2$ parameters. The proposed Ising similarity regression model is thus particularly useful when the dimension of the binary response vector $p$ is large: even when the number of similarity matrices $K$ grows at the same rate as the dimension $p$, the number of parameters in model \eqref{eq:ising_regression_model} only grows linearly in $p$ compared to $O(p^2)$ parameters in model \eqref{eq:ising_regression_model0}.

\subsection{Estimation and  \zy{Similarity} Selection}\label{subsec:estimation}

Suppose we have observations of $p$-dimensional response vectors $\bm{y}_i = (y_{i1},\cdots, \\ y_{ip})^\top \in \{0,1\}^{p}$ for $i=1,\cdots,n$,  where $\bm{y}_1,\cdots,\bm{y}_n\stackrel{i.i.d.}\sim f(\cdot;\bmvartheta)$ follows the Ising similarity regression model in equation \eqref{eq:ising_regression_model}, and $i.i.d.$ denotes independent and identically distributed. Recall the pmf $f(\cdot;\bmvartheta)$ involves the normalization constant $Z(\bmvartheta)$, which is a sum of $2^p$ terms. As a result, maximum likelihood estimation based on $\prod_{i=1}^n f(\bm{y}_i;\bmvartheta)$ is computationally not feasible when the dimension $p$ is large.  To overcome this problem, we adapt the existing literature (e.g., \citeauthor{hoflingANDtibshirani2009}, \citeyear{hoflingANDtibshirani2009}; \citeauthor*{ravikumarETAL2010}, \citeyear{ravikumarETAL2010}) and propose a pseudo-likelihood estimation approach to fit the Ising similarity regression model. Let $\bm{y}_{i\backslash j} = (y_{i1},\cdots,y_{i,j-1},  y_{i,j+1}, \cdots,y_{ip})^\top$ denote the $i$-th observed response vector without the $j$-th element, and $f_j(\cdot | \bm{y}_{i\backslash j}; \bmvartheta)$ denote the corresponding conditional pmf of $y_{ij}$ given $\bm{y}_{i\backslash j}$ for $i=1,\cdots,n$ and $j=1,\cdots,p$. Then the (unregularized) pseudo-likelihood  estimator can be obtained by maximizing the pseudo-likelihood function  $\prod_{i=1}^{n} \prod_{j=1}^{p} f_j( y_{ij} | \bm{y}_{i \backslash j}; \bmvartheta )$. In particular, the conditional pmf $f_j(\cdot | \bm{y}_{i\backslash j}; \bmvartheta)$ can be derived based on $f(\cdot;\bmvartheta)$ in equation \eqref{eq:ising_regression_model}, and takes the simple form below,
\begin{equation}
	    f_j(u | \bm{y}_{i\backslash j}; \bmvartheta) 
	    = \frac{\exp \lbk u \lsk \theta_{jj} + \sum_{k=1}^{K} \alpha_k \sum_{j' \neq j} w_{jj'}^{(k)}y_{ij'} \rsk \rbk}{1 + \exp \lsk \theta_{jj} + \sum_{k=1}^{K} \alpha_k \sum_{j' \neq j} w_{jj'}^{(k)}y_{ij'} \rsk},
	    \label{eq:conditional_pmf}
\end{equation}
for $u \in \{0,1\}$, $i=1,\cdots,n$ and $j=1,\cdots,p$. It follows that the conditional log-odds is given by
\begin{equation}
	    \log \lbk \frac{ \mathrm{P}(y_{ij} = 1 | \bm{y}_{i\backslash j} ; \bmvartheta)}{1 -\mathrm{P}(y_{ij} = 1 | \bm{y}_{i\backslash j} ; \bmvartheta) } \rbk = \theta_{jj} + \sum_{k=1}^{K} \alpha_k \sum_{j' \neq j}w_{jj'}^{(k)} y_{ij'},
		\label{eq:log_odds}
\end{equation}
for $i=1,\cdots,n$ and $j=1,\cdots,p$, from which we observe that the conditional pmf $f_j(\cdot | \bm{y}_{i \backslash j}; \bmvartheta)$ and thus pseudo-likelihood estimation no longer involve the intractable normalization constant $Z(\bmvartheta)$. In fact, equation \eqref{eq:log_odds} bears a similar form to  fitting a logistic regression for the conditional log-odds of  $y_{ij} = 1$ against the set of $K$ covariates $\{\sum_{j' \neq j} w_{jj'}^{(k)} y_{ij'}=\bm{W}_{j\cdot}^{(k)\top}\bm{y}_{i}:k=1,\cdots,K\}$, with an intercept term $\theta_{jj}$ and regression coefficients $\bmalpha = (\alpha_1,\cdots, \alpha_K)^\top$, where $\bm{W}_{j\cdot}^{(k)\top}\in\mathbb{R}^{1\times p}$ denotes the $j$-th row of $\bmW_k$. 

We can further augment the above pseudo-likelihood estimator with a penalty to perform variable selection on the elements of $\bmalpha$. 
This is useful in practice when there are a non-negligible number of similarity matrices available for data analysis, but only a subset of them are anticipated to be relevant in equation \eqref{eq:matrix_regression}. To perform simultaneous estimation and regularization on the coefficient vector $\bmalpha$, we augment the log pseudo-likelihood function based on equation \eqref{eq:conditional_pmf} with an adaptive lasso penalty \citep{zou2006}, resulting in a regularized pseudo-likelihood estimator that minimizes the objective function 
\begin{align}
	&- \frac{1}{np} \sum_{i = 1}^{n} \sum_{j=1}^{p} \log \{ f_j( y_{ij} | \bm{y}_{i \backslash j}; \bmvartheta ) \} +  \lambda \sum_{k=1}^{K} w_k \lak \alpha_k \rak \notag \\
	=&- \frac{1}{np} \sum_{i=1}^{n} \sum_{j=1}^{p}\left[ y_{ij} \lsk \theta_{jj} + \sum_{k=1}^{K} \alpha_k \bm{W}_{j\cdot}^{(k)\top}\bm{y}_{i} \rsk \right. \notag \\
        & - \left. \log  \lbk 1+ \exp \lsk \theta_{jj}+ \sum_{k=1}^{K} \alpha_k \bm{W}_{j\cdot}^{(k)\top}\bm{y}_{i} \rsk \rbk \right] +  \lambda \sum_{k=1}^{K} w_k \lak \alpha_k \rak,
		\label{eq:penalized_pseudo_log_likelihood}
\end{align}
given a tuning parameter $\lambda > 0$. Following \citet{zou2006} and \citet*{huang2008adaptive} among others, we set the adaptive weights as $w_k = 1 / |\bar{\alpha}_k|$ for $k=1,\cdots,K$, where $ \bar \bmalpha = (\bar \alpha_1 , \cdots, \bar \alpha_K)^\top $ denotes the unregularized pseudo-likelihood  estimator i.e., the estimator of $\bmalpha$ which minimizes the objective function \eqref{eq:penalized_pseudo_log_likelihood} with $\lambda = 0$. It is worth noting that $\theta_{jj}$, the $j$-specific intercepts for $j=1,\cdots,p$, are not  regularized; this is similar to the literature of sparse Ising models (\citeauthor{guoETAL2010}, \citeyear{guoETAL2010}; \citeauthor*{ravikumarETAL2010}, \citeyear{ravikumarETAL2010}) as well as other regularized regression settings in general \citep*[e.g,][]{hui2017joint,hui2018sparse}. Regarding the choice of the penalty function, in this article we focus our developments on the adaptive lasso  penalty $\lambda \sum_{k=1}^{K} w_k \lak \alpha_k \rak$; see Section \ref{sec:conclusion} for a discussion on alternative penalty functions. In particular, with adaptive weights there is only a single regularization parameter $\lambda$ in the penalty, and the whole objective function \eqref{eq:penalized_pseudo_log_likelihood} remains convex for optimization. Moreover, the incorporation of adaptive weights $w_k$ allows for varying degrees of regularization on the coefficients $\alpha_k$ across $k=1,\cdots,K$, and facilitates selection consistency for a sparse coefficient vector $\bmalpha$ which we theoretically examine in the next section. \zy{\label{page:similarity_selection_section2}We also emphasize that the adaptive lasso penalty here is used for the selection of regression coefficients associated with similarity matrices i.e., similarity selection, and not for the selection of edges in $\bm{\Theta}$. This is apparent when we see that the penalty induces sparsity in the regression coefficients, but not the similarity matrices and the resulting $\bm{\Theta}$.}
The details of implementing the regularized pseudo-likelihood estimation by minimizing \eqref{eq:penalized_pseudo_log_likelihood} are discussed in Section \ref{sec:simulation}.
 
\section{Theoretical Results} \label{sec:theory}
In this section, we establish asymptotic properties for the regularized pseudo-likelihood estimator of the Ising similarity regression model \eqref{eq:ising_regression_model}. Since our main interest lies in estimating similarity regression coefficients $\bmalpha$ that only depend on the pairwise interaction coefficients in equation \eqref{eq:model1}, then we focus on a variant of the model with no main effects, giving rise to the criterion
\begin{equation}\label{eq:cri1} 
\hat{\bmalpha} = \arg\min_{\bm{\alpha}} \lbk -l(\bm{\alpha}) + \lambda\sum_{k=1}^{K} w_k \lak \alpha_k \rak\rbk,
\end{equation}
where $ l(\bmalpha) = \sum_{i=1}^{n} l_i (\bmalpha) / (np) $, $l_i(\bmalpha) = \sum_{j=1}^{p}[ y_{ij} (  \sum_{k=1}^{K} \alpha_k \bm{W}_{j\cdot}^{(k)\top}\bm{y}_{i} ) - \log \{ 1+ \exp ( \sum_{k=1}^{K} \alpha_k $ $\bm{W}_{j\cdot}^{(k)\top}\bm{y}_{i} ) \} ]$, $w_k = 1/|\bar{\alpha}_k|$ and $\bar{\bm{\alpha}} = (\bar{\alpha}_1,\cdots,\bar{\alpha}_K)^\top = \arg\min_{\bmalpha} \{ \\ -l(\bmalpha)\}$; see  \cite{guoETAL2010, guoETAL2015} for a similar theoretical treatment. \zy{\label{page:theory_emphasize_Ising}It is worth noting that while the pseudo-likelihood function is used in criterion \eqref{eq:cri1} to overcome the issue of intractable normalization constant, theoretical results in this section are still obtained under the joint probability distribution of the Ising similarity regression model.}

Let $ \bmalphao = (\alphao_1,\cdots,\alphao_K)^\top $ denote the true value of the coefficient vector $ \bmalpha $, $ S = \{k : \alphao_k \neq 0,\textrm{ for }k=1,\cdots,K\} $ denote the set indexing all truly non-zero coefficients, $S^c=\{1,\cdots,K\}\backslash S$, and let the cardinality of $S$ be denoted by $ |S| = K_0$, where $ K_0 $ is assumed to be finite. Furthermore, we construct
   \begin{equation}
		\bmmX^{(i)} = 
		\begin{pmatrix}
		    \bmmX^{(i,1)\top} \\
		    \vdots \\
		    \bmmX^{(i,p)\top}
		\end{pmatrix} 
      =   \begin{pmatrix}
			\bm{W}_{1\cdot}^{(1)\top} \bm{y}_i & \cdots & \bm{W}_{1\cdot}^{(K)\top} \bm{y}_i \\
			\vdots & \ddots & \vdots \\
			\bm{W}_{p\cdot}^{(1)\top} \bm{y}_i & \cdots & \bm{W}_{p\cdot}^{(K)\top} \bm{y}_i
		\end{pmatrix}, \textrm{ for } i=1,\cdots,n,
		\label{eq:bmmX}
	\end{equation}
and define the $K\times K$ matrices $ \bm{U}^0 = E\{\sum_{i=1}^{n} \bmmX^{(i)\top} \bmmX^{(i)} / (np) \}$ and $ \bm{M}^0 = E \{-\nabla^2 l(\bmalphao)\} $. Finally, for a generic $m\times r$ matrix $\bm{H} = (h_{tv})_{m \times r}$, and subsets of row and column indices $\mathcal{T}\subseteq\{1,\cdots,m\}$ and $\mathcal{V}\subseteq\{1,\cdots,r\}$, let $\bm{H}_{\mathcal{T},\mathcal{V}}$ denote the submatrix of $\bm{H}$ consisting of rows and columns indexed by $\mathcal{T}$ and $\mathcal{V}$, respectively, $ \laak \bm{H} \raak_1 = \max_{1 \leq v \leq r} \{ \sum_{t=1}^{m} |h_{tv}| \} $ denote the matrix 1-norm of $ \bm{H} $, and $\Lambda_{\min}(\bm{H})$ and $\Lambda_{\max}(\bm{H})$ denote the smallest and largest eigenvalues of $\bm{H}$, respectively, when $\bm{H}$ is a square matrix (i.e., $m=r$). \zy{\label{page:notation}We also refer the reader to Table \ref{table:notation} in Section \ref{sec:notation} of the supplementary material for a list of important notations used throughout this article along with their definitions.}
	
We introduce the following technical conditions.
	\begin{condition}
		There exist finite positive constants $C_{\min}$ and $C_{\max}$ such that $\Lambda_{\min}(\bm{M}^{0}) \\ \geq C_{\min}$ and $\Lambda_{\max}(\bm{U}^{0}) \leq  C_{\max}$\zy{, where $\bm{M}^0$ and $\bm{U}^0$ are defined below equation \eqref{eq:bmmX}.}
		\label{cond:eigenvalue_pop}
	\end{condition}
	
	\begin{condition}
		There exists a finite positive constant $C_W$ such that $\laak \bmW_k \raak_1 \leq C_W$ for all symmetric similarity matrices $\{\bmW_k: k=1,\cdots,K\}$. 
		\label{cond:W}
	\end{condition}

	By the Cauchy Interlacing Theorem \citep[][p. 186]{parlett1980} and Condition \ref{cond:eigenvalue_pop}, similar conditions  hold for the submatrices $\bm{M}_{S,S}^{0}$ and $\bm{U}_{S,S}^{0}$; in particular, $\Lambda_{\min}(\bm{M}_{S,S}^{0}) \geq C_{\min}$ and $\Lambda_{\max}(\bm{U}_{S,S}^{0}) \leq C_{\max}$. Condition \ref{cond:W} is \zy{\label{page:example_cond_W_start}a bounded matrix norm assumption which is} similar to the conditions imposed on the similarity matrices $ \bmW_k $  in the literature \citep[see for instance, Condition C8 in][]{tao2021}. \zy{For example, when $\bm{W}_k$ is a symmetric adjacency matrix with elements $w_{jj'}^{(k)} \in \{0,1\}$ capturing the neighborhood relationship among $p$ nodes, the condition is equivalent to a column-sparsity (and row-sparsity since $\bm{W}_k$ are symmetric) condition that requires the number of neighbors for each node to be finite even when the total number of nodes $p$ diverges.\label{page:example_cond_W_end} Other classes of similarity matrices such as those with bounded elements $|w_{jj'}^{(k)}| \leq C $ and $s$-sparse column (and row) vectors \citep[e.g.,][p. 156]{wainwright2019} i.e., $\sum_{j=1}^{p} 1_{\{ |w_{jj'}^{(k)}| > 0 \}} \leq s $, where $C$ and $s$ are finite positive constants and  $1_{\{\cdot\}}$ is the indicator function, also satisfy Condition \ref{cond:W} with $C_W = sC$. \label{page:bounded_matrix_norm_Theta0}Together with the above assumed sparsity on the true coefficient vector $\bmalphao$, Condition \ref{cond:W} implies similar bounded matrix norm condition for the true interaction matrix $\bm{\Theta}^{(0)} = \sum_{k=1}^{K} \alphao_k \bm{W}_k$; that is, $\|\bm{\Theta}^{(0)}\|_1 \leq C_W \sum_{k \in S} |\alphao_{k} | $.}
    
    We first establish the estimation consistency for the unregularized pseudo-likelihood estimator that is used to construct the adaptive weights.
	\begin{theorem}
		Assume Conditions \ref{cond:eigenvalue_pop} -- \ref{cond:W} are satisfied. If $K\sqrt{\log(p)/n}=o(1)$ and there exists a finite positive constant $C_\nabla$ such that $K = o ( p^{ C_{\nabla}^2/( 8 C_W^2)} )$  as $n,p \to\infty$, then with probability tending to one it holds that 
		$
			\| \bar \bmalpha - \bmalphao \|_2 \leq \bar M \sqrt{K \log(p) / n}, \notag 
		$
		for a finite positive constant $\bar M > 4 C_{\nabla} /  C_{\min} $.
  \label{THM:CONSISTENT_UNPENALIZED}
	\end{theorem}
	The proofs of Theorem \ref{THM:CONSISTENT_UNPENALIZED} along with all other theoretical results are provided in the supplementary material. The above convergence rate bears a similar form to the convergence rate of many lasso-penalized estimators \citep[e.g.,][]{guoETAL2010,nghiem2022sparse}, where the factor of $K$, instead of $K_0$, in the numerator is expected given this is the unregularized estimator. In addition, Theorem \ref{THM:CONSISTENT_UNPENALIZED} is valid under both cases of fixed $K$ and diverging $K$, so long as $K \sqrt{\log(p)/n} \to 0$ and $K / p^{ C_{\nabla}^2/( 8 C_W^2)} \to 0 $.
 	
    To study the asymptotic properties of the regularized pseudo-likelihood estimator, we introduce two additional technical conditions.

	\begin{condition}
		There exists a constant $C_M \in (0,1)$ such that $\| \bm{M}_{S^c,S}^{0} (\bm{M}_{S,S}^{0})^{-1}  \|_\infty \leq 1 - C_M$.
		\label{cond:irrepresent_pop}
	\end{condition}

	\begin{condition}
		$\lambda \sqrt{n} / \{ \min\{|\alphao_k| : k \in S\} \sqrt{\log(p)} \} \to 0$ and $\lambda n / \{\sqrt{K} \log(p)\} \to \infty$.
		\label{cond:lambda}	
	\end{condition}
	
	Condition \ref{cond:irrepresent_pop} is commonly known as the mutual incoherence or irrepresentability condition \citep*{hastieETAL2015}, and, together with the conditions on $\bm{M}_{S,S}^{0}$ and $\bm{U}_{S,S}^{0}$ implied by Condition \ref{cond:eigenvalue_pop}, are similar to those assumed by \cite{meinhausenANDbuhlmann2006} and \cite{guoETAL2010} among others. Condition \ref{cond:lambda} is similar to existing conditions in the literature regarding the rates of the tuning parameters for adaptive lasso regression \citep*[e.g.,][]{wang2009shrinkage,hui2018sparse}. 
 
 We now state the main results of this paper for the regularized pseudo-likelihood estimator $\hat{\bmalpha}  = (\hat{\alpha}_1,\cdots,\hat{\alpha}_K)^\top$ of the Ising similarity regression model.
	\begin{theorem}
		Assume Conditions \ref{cond:eigenvalue_pop} -- \ref{cond:lambda} are satisfied. If $K\sqrt{\log(p)/n}=o(1)$ and  there exists a finite positive constant $C_\nabla$ such that $K = o ( p^{ C_{\nabla}^2/( 8 C_W^2)} )$ as $n,p \to\infty$, then with probability tending to one it holds that
		\begin{enumerate}[(a)]
			\item $\laak \hat{\bmalpha} - \bmalphao \raak_2 \leq M \sqrt{K_0 \log(p)/n}$ for a finite positive constant $M > 4/C_{\min}$;
			\item $\hat{\alpha}_k \neq 0$ for all $k \in S$ and $ \hat{\alpha}_k = 0$ for all $k \in S^c$.
		\end{enumerate}
  \label{THM:CONSISTENT_ADALASSO}
	\end{theorem}
	
	Theorem \ref{THM:CONSISTENT_ADALASSO}(a) establishes estimation consistency of the regularized pseudo-likelihood estimator, noting its convergence rate can be faster than that of the unregularized estimator. Theorem \ref{THM:CONSISTENT_ADALASSO}(b) establishes selection consistency of the regularized pseudo-likelihood estimator i.e., it can recover the underlying sparsity pattern of $\bm{\alpha}^{(0)}$. This is attractive when the Ising similarity regression model is applied to datasets where the dimension of responses and number of similarity matrices may be large. 

To conclude this section, we remark Condition \ref{cond:lambda} provides requirements on the rate of the tuning parameter $\lambda$ based on the smallest truly non-zero coefficient $\min\{ | \alphao_k |: k \in S\}$, as the basis for establishing the results in Theorem \ref{THM:CONSISTENT_ADALASSO}. For instance, if $\min\{ | \alphao_k | : k \in S\}$ is bounded away from zero, then it can be verified that $\lambda = \{n / \log(p)\}^{-t} K^{v}$ for some $t \in (1/2,1)$ and $v \in [2t-3/2, 2t-1]$ suffices for Condition \ref{cond:lambda} to hold. If this is relaxed and we permit $\min\{ |\alphao_k | : k \in S\}$ to tend to zero at a rate satisfying $\min\{ | \alphao_k | : k \in S\} \{n/ \log(p)\}^{m} \to \infty$ for some $m \in (0,1/4]$, then  $\lambda = O(\{\log(p)/n\}^{q})$ for some $q \in [m+1/2, 3/4]$ will satisfy the requirements. In practice, since $\min\{ | \alphao_k | : k \in S\}$ is unknown, \zy{\label{page:typo1}we adopt a data-driven approach to select $\lambda$ as discussed in the next section.}

\section{Simulation Study} \label{sec:simulation}
We conduct a numerical study to evaluate the finite sample performance of the proposed regularized pseudo-likelihood estimator for the Ising similarity regression model \eqref{eq:ising_regression_model}. Briefly, we consider sample sizes $ n \in \{50,100,200,400\} $, numbers of binary responses $ p \in \{10,25,50,100,200\} $, and $K = 20$ similarity matrices with only the first $K_0 = 5$ of them having non-zero true regression coefficients i.e., $S = \{1,\cdots,5\}$. A total of 1000 datasets are simulated from the Ising similarity regression model \eqref{eq:ising_regression_model} for each combination of $n$ and $p$. Details of the simulation settings, including the true values of the main effect parameters  $\{\theta_{jj}^{(0)}: j = 1,\cdots,p\}$ and regression coefficients $\{\alphao_k: k=1,\cdots,K\}$ are provided in Section \ref{sec:simulation_supp} of the supplementary material.
 
We employ a two-step algorithm to compute the regularized pseudo-likelihood estimator. First, from the discussion below equations \eqref{eq:log_odds} -- \eqref{eq:penalized_pseudo_log_likelihood}, the unregularized estimator is obtained by fitting a logistic regression model with $\bm{y}_{1:n} = (y_{11},\cdots, y_{n1},\cdots, y_{1p},\cdots, y_{np})^\top$ as the response and $(\bm{I}_p \otimes \bm{1}_n, \bmmX)$ as the model matrix, where $\bm{1}_n$ is an $n$-dimensional vector of ones, $\otimes$ is the Kronecker product operator, \zy{\label{page:bmmX}$\bmmX = ( \bmmX^{(1,1)}, \cdots, \bmmX^{(n,1)}, \cdots, \bmmX^{(1,p)}, \cdots, \bmmX^{(n,p)})^\top$, and $\bmmX^{(i,j)}$ denotes the $j$-th row of the matrix $\bmmX^{(i)}$ given in equation \eqref{eq:bmmX}.} In the second step, we compute the regularized pseudo-likelihood estimator by fitting an adaptive lasso regularized logistic regression to $\bm{y}_{1:n}$ and $(\bm{I}_p \otimes \bm{1}_n, \bmmX)$ via the \texttt{R} package \texttt{glmnet} \citep*{glmnetpackage}, where the adaptive regularization weights in  \eqref{eq:penalized_pseudo_log_likelihood} are constructed based on the unregularized estimator.
	
To select the tuning parameter $\lambda$ in \eqref{eq:penalized_pseudo_log_likelihood}, we utilize a ten-fold cross-validation approach, where the observations in each dataset are grouped at the level indexed by $ i = 1,\cdots,n $. That is, we split the data $ \{(\bm{y}_1, \bmmX^{(1)}), \cdots, (\bm{y}_n, \bmmX^{(n)})\} $ randomly into ten folds and select an optimal $\lambda$ that gives the largest mean pseudo-likelihood (averaged across ten test sets), noting that the same set of adaptive weights constructed based on the unregularized estimator from the full dataset is used throughout the cross-validation. The grouping of the observations is done to preserve the dependence structure of the data, since $y_{ij}$ are independent in the index $i$ but dependent in the response index $j$ through the Ising model \citep*[see also][]{warton2017pit}. 
	
We compare our proposed regularized pseudo-likelihood estimator (Regularized) with three basic approaches: \zy{\label{page:lasso_estimator}the lasso-regularized estimator (Lasso) based on \eqref{eq:penalized_pseudo_log_likelihood} with $w_k = 1$ for $k=1,\cdots, K$ and $\lambda$ selected using a similar ten-fold cross-validation approach,} the unregularized estimator (Unregularized) based on  \eqref{eq:penalized_pseudo_log_likelihood} with $\lambda = 0$, and the oracle estimator (Oracle) based on minimizing an alternative to \eqref{eq:penalized_pseudo_log_likelihood} where $\lambda = 0$ and $\alpha_k$ are set to be zeros for $k \in S^c$. \zy{The lasso-regularized estimator is considered to study the effect of penalty choice on the empirical performance of the estimator,} the unregularized estimator is included to examine whether regularization could lead to better overall estimation performance, and the oracle estimator is used as a benchmark since it incorporates additional information regarding the index set $S = \{1,\cdots,5\}$. \zy{We compare point estimation performance for all four estimators of the regression coefficients and main effect parameters using their mean square error (MSE), $\textrm{MSE}_{\alpha}$ and $\textrm{MSE}_{\theta}$, respectively, and the model selection performance of the proposed estimator and lasso-regularized estimator of the regression coefficients using the true positive rate (TPR) and false positive rate (FPR); see Section \ref{sec:simulation_supp} of the supplementary material for details on the computation of these performance measures.}

\begin{table}[t!]                               
    \caption{MSE for the Oracle, Regularized, Lasso, and Unregularized estimators of the regression coefficients ($\textrm{MSE}_\alpha$) and main effect parameters ($\textrm{MSE}_\theta$) in the simulation study involving the Ising similarity regression model. $\textrm{MSE}_\alpha$ is multiplied by 1000 for clarity. \label{table:simulation_mse}} \centering \medskip
	   \resizebox{\linewidth}{!}{\begin{tabular}{|l l  r r r r r r r r r|}             			\hline
			& & \multicolumn{4}{c}{$1000 \times \textrm{MSE}_\alpha$} & & \multicolumn{4}{c|}{$\textrm{MSE}_\theta$} \\ \hline
			$p$ & $n$ & Oracle & Regularized & Lasso & Unregularized & & Oracle & Regularized & Lasso & Unregularized  \\[3pt] \hline
    \multirow{4}{*}{10} & 50 & 45.791 & 68.420 & 37.524 & 351.622 & & 0.350 & 0.402 & 0.346 & 0.896\\
     & 100 & 21.544 & 44.121 & 28.569 & 147.225 & & 0.167 & 0.207 & 0.193 & 0.370\\
     & 200 & 10.755 & 27.506 & 19.048 & 68.262 & & 0.082 & 0.111 & 0.099 & 0.172\\
     & 400 & 5.390 & 16.298 & 11.717 & 33.021 & & 0.038 & 0.057 & 0.052 & 0.080\\
    \hline
    \multirow{4}{*}{25} & 50 & 5.858 & 14.717 & 13.474 & 25.985 & & 0.298 & 0.391 & 0.351 & 0.546\\
     & 100 & 2.846 & 6.817 & 6.992 & 12.089 & & 0.137 & 0.184 & 0.171 & 0.242\\
     & 200 & 1.434 & 3.025 & 3.546 & 5.996 & & 0.068 & 0.088 & 0.087 & 0.118\\
     & 400 & 0.710 & 1.356 & 1.780 & 2.950 & & 0.033 & 0.041 & 0.043 & 0.057\\
    \hline
    \multirow{4}{*}{50} & 50 & 2.100 & 5.901 & 5.908 & 8.368 & & 0.550 & 0.788 & 0.719 & 0.737\\
     & 100 & 0.969 & 1.924 & 2.537 & 3.967 & & 0.120 & 0.156 & 0.166 & 0.205\\
     & 200 & 0.469 & 0.860 & 1.266 & 1.935 & & 0.053 & 0.067 & 0.074 & 0.093\\
     & 400 & 0.232 & 0.386 & 0.625 & 0.939 & & 0.027 & 0.033 & 0.038 & 0.047\\
    \hline
    \multirow{4}{*}{100} & 50 & 0.765 & 4.517 & 3.282 & 3.065 & & 2.314 & 1.551 & 2.201 & 2.444\\
     & 100 & 0.362 & 1.390 & 1.609 & 1.441 & & 0.500 & 0.559 & 0.512 & 0.567\\
     & 200 & 0.175 & 0.367 & 0.510 & 0.709 & & 0.099 & 0.111 & 0.113 & 0.132\\
     & 400 & 0.089 & 0.119 & 0.227 & 0.355 & & 0.033 & 0.034 & 0.041 & 0.049\\
    \hline
    \multirow{4}{*}{200} & 50 & 0.351 & 5.493 & 2.406 & 1.282 & & 1.971 & 2.180 & 2.420 & 2.073\\
     & 100 & 0.169 & 0.376 & 0.422 & 0.614 & & 0.290 & 0.350 & 0.395 & 0.343\\
     & 200 & 0.079 & 0.108 & 0.197 & 0.295 & & 0.073 & 0.076 & 0.089 & 0.102\\
     & 400 & 0.039 & 0.061 & 0.096 & 0.148 & & 0.030 & 0.031 & 0.035 & 0.044\\
			\hline
		\end{tabular}}   \vspace{-1em}   
\end{table}  

Table \ref{table:simulation_mse} shows the $\textrm{MSE}_{\alpha}$ for all four estimators decrease as $ n $ and $ p $ increase, while their $\textrm{MSE}_{\theta}$ exhibit a clear decreasing trend when $ n $ increases but no obvious patterns when $p$ increases; this is to be expected given the main effect parameter $ \theta_{jj} $ is specific to each of the responses in the Ising similarity regression model for $j=1,\cdots,p$. The $\textrm{MSE}_{\alpha}$ and $\textrm{MSE}_{\theta}$ for the proposed regularized pseudo-likelihood estimators are typically much smaller than the unregularized estimators, and in fact similar to the oracle estimators especially when $n$ and $p$ are large. This is consistent with the notion that the proposed estimators take advantage of the underlying sparsity in the model, resulting in better overall estimation performance than the unregularized estimators. \zy{\label{page:lasso_mse_discussion}Furthermore, the MSEs of the proposed estimators tend to be smaller than the lasso-regularized estimators when $n$ and $p$ are large.} 
	
\begin{table}[t!]                                 \caption{TPR and FPR for the Regularized (top panel) and Lasso (bottom panel) estimators of the regression coefficients in the simulation study involving the Ising similarity regression model.\label{table:simulation_tpr_fpr}} \centering \medskip
	\resizebox{0.73\linewidth}{!}{\begin{tabular}{|l  r r r  r r r r r  r|}                                   \hline
        \multicolumn{10}{|c|}{Regularized} \\
			& \multicolumn{4}{c}{TPR} & & \multicolumn{4}{c|}{FPR} \\ \hline
			$p \backslash n$ & 50 & 100 & 200 & 400 & & 50 & 100 & 200 & 400  \\[3pt] \hline  
			10 & 0.337 & 0.490 & 0.654 & 0.824 & & 0.167 & 0.202 & 0.225 & 0.235 \\ 
            25 & 0.835 & 0.970 & 0.999 & 1 & &0.218 & 0.218 & 0.197 & 0.167 \\ 
            50 & 0.954 & 0.999 & 1 & 1 & &0.186 & 0.174 & 0.154 & 0.102 \\ 
            100 & 0.976 & 0.998 & 0.999 & 1 & &0.043 & 0.105 & 0.076 & 0.009 \\ 
            200 & 0.925 & 0.998 & 1 & 1 & &0.043 & 0.058 & 0.005 & 0 \\ 
			\hline
            \multicolumn{10}{|c|}{Lasso} \\
            & \multicolumn{4}{c}{TPR} & & \multicolumn{4}{c|}{FPR} \\ \hline
			$p \backslash n$ & 50 & 100 & 200 & 400 & & 50 & 100 & 200 & 400  \\[3pt] \hline  
			10 & 0.307 & 0.500 & 0.711 & 0.888 &  & 0.124 & 0.185 & 0.237 & 0.290\\
            25 & 0.859 & 0.980 & 1 & 1 &  & 0.298 & 0.356 & 0.392 & 0.395\\
            50 & 0.982 & 0.999 & 1 & 1 &  & 0.403 & 0.412 & 0.426 & 0.428\\
            100 & 0.999 & 1 & 1 & 1 &  & 0.294 & 0.327 & 0.405 & 0.429\\
            200 & 0.996 & 1 & 1 & 1 &  & 0.254 & 0.354 & 0.392 & 0.433\\ 
			\hline
		\end{tabular}} \vspace{-1em}
\end{table}

Turning to the model selection performance of the proposed estimator in Table \ref{table:simulation_tpr_fpr}, while there is slight underfitting as reflected by the comparably low TPR for the case of smallest $p = 10$, both the TPR and FPR tend to one and zero, respectively, as $ n $ and $ p $ increase. This is consistent with Theorem \ref{THM:CONSISTENT_ADALASSO}(b), empirically demonstrating that the proposed method is able to recover the underlying sparsity in the Ising similarity regression model. \zy{\label{page:lasso_tpr_fpr_discussion}The lasso-regularized estimator suffers from overfitting as seen from its FPR not converging towards zero, thus supporting the use of the adaptive lasso penalty in our proposed estimator. 
    
\label{page:aic_bic_discussion}Results from comparing our ten-fold cross-validation approach to the use of AIC and BIC criteria for choosing the tuning parameter are provided in Table \ref{table:simulation_tpr_fpr_aic_bic} of the supplementary material. Overall, results show the BIC performs similarly well as our cross-validation approach, while the AIC suffers from clear overfitting when $p$ is small due to its weaker model complexity penalty. These results provide empirical support for using the cross-validation approach to choose $\lambda$. 

\label{page:traditional_discussion_start}Next, we compare the proposed estimator to other traditional estimators of the Ising model in the literature that involve direct estimation of the Ising model interaction matrix $\bm{\Theta}$, without considering the similarity matrices $\bm{W}_k$ \citep[e.g., Section 3 of][]{hoflingANDtibshirani2009}. From Table \ref{table:simulation_fsnorm} in the supplementary material, we see that the proposed estimator greatly outperforms the traditional Ising model estimators in recovering the true $\bm{\Theta}$ matrix, since it incorporates the additional information from the similarity measures.\label{page:traditional_discussion}

Finally, we conduct simulation studies with varying number of similarity matrices $K \in \{10,20,40,80,200\}$ while keeping the same number of truly non-zero regression coefficients $K_0 = 5$, as a further investigation on the effect of $K$ on the empirical performance of various estimators. Unsurprisingly, the estimation and model selection performance of all methods decline as more irrelevant similarity matrices are being added, noting the proposed estimator still performs reasonably well relative to other estimators under different settings of $K$; see Tables \ref{table:simulation_mse_varying_K} -- \ref{table:simulation_fsnorm_varying_K} in the supplementary material for full results. \label{page:varying_K_discussion}} 

 \section{Application to U.S. Senate Roll Call Voting Data} \label{sec:real_data}
We apply the Ising similarity regression model \eqref{eq:ising_regression_model} to roll call voting data from the U.S. Senate as part of the 117-th Congress, covering the period from 6 January 2021 to 20 May 2021 (date of data collection). Roll call voting data has previously been studied using a variety of statistical techniques \zy{\label{page:typo2}including undirected graphical models \citep*{banerjeeETAL2008}.} Here, we use our model to study how voting associations between senators are associated with their similarities in various demographic attributes and social network profiles. The dataset is obtained from the U.S. Senate's website (\url{https://www.senate.gov/}), which originally consists of binary voting records, coded as one for `Yea' and zero for `Nay', on 199 bills by 100 senators of the 117-th Congress. After performing some preliminary data wrangling procedures (see Section \ref{sec:additional_results_rollcall} of the supplementary material), the final dataset analyzed consists of $n = 138$ bills voted by $p = 100$ senators, where all bills are treated as independent (e.g., \citeauthor*{banerjeeETAL2008}, \citeyear{banerjeeETAL2008}; \citeauthor{guoETAL2010}, \citeyear{guoETAL2010}).
 
We obtain several attributes for each senator, including their state, political party, class, and age from the U.S. Senate website, along with gender and occupation from Wikipedia. Additionally, each senator's Twitter handle is obtained from the `us-senate' GitHub project of `CivilServiceUSA', and used to compute the number of tweets and number of followers for each senator. \zy{Each of these attributes is then converted into a similarity matrix $\bm{W}_k$ following the procedure described in Section \ref{subsec:model_setup}, depending on whether it is a qualitative (state, party, class, gender and occupation) or quantitative (age, number of tweets, number of Twitter followers) attribute.\label{page:refer_description_analysis_senator} We refer the reader to Section \ref{sec:additional_results_rollcall} of the supplementary material for detailed construction of these similarity matrices, as well as a descriptive analysis for the above attributes and the binary votes of the senators. \label{page:refer_description_analysis_senator_end} Note that a symmetric adjacency matrix $\bm{W}_k$ is also constructed to summarize the Twitter follower-followee relationship among the senators, such that $ w_{jj'}^{(k)} = 1 $ if the $ j $-th senator follows the $ j' $-th senator on Twitter or vice versa, and zero otherwise. As a result, the analysis consists of $ K  = 15$ similarity matrices.} 
    
We fit the Ising similarity regression model using the regularized pseudo-likelihood estimator to identify and quantify truly important attributes driving the voting associations between senators, where $\lambda$ is selected using ten-fold cross-validation with the groupings being done at the bill level. After performing variable selection and obtaining the estimated non-zero regression coefficients, we construct 95\% Wald confidence intervals for each coefficient based on standard errors from the empirical sandwich covariance matrix\zy{\label{page:wald_discussion} obtained by deriving the score and Hessian matrix of the log pseudo-likelihood for each observation with respect to the set of chosen similarity measures; see Section \ref{sec:inference} of the supplementary material for details of its derivation.}

Table \ref{table:rollcall_vote} shows only 7 of the 15 regression coefficients are estimated to be non-zero using the adaptive lasso penalty; estimation results for the main effect parameters are given in the supplementary material. Of these, unsurprisingly there is statistically clear evidence that senators from the same state and/or party are more likely to vote similarly on the bills (analogous state and party effects are found in \citeauthor*{banerjeeETAL2008}, \citeyear{banerjeeETAL2008}; \citeauthor{guoETAL2010}, \citeyear{guoETAL2010}, among others), although the former exhibits a much stronger effect. Most occupations are found to be uninformative for the conditional dependence relationships between senator voting patterns, except for businessman and lawyer, although the confidence interval corresponding to the effect of lawyer contains zero. The presence of a positive effect for the businessman similarity matrix could be attributed to senators who are businessmen tending to vote similarly on bills related to the economy. 
	
\begin{table}[t!]                                
 \caption{Point estimates and 95\% confidence intervals (in parentheses) for the regression coefficients corresponding to the $K = 15$ similarity matrices, based on fitting the Ising similarity regression model \eqref{eq:ising_regression_model} to the U.S. Senate roll call voting data using regularized pseudo-likelihood estimation. Estimates whose corresponding confidence interval excludes zero are bolded. \label{table:rollcall_vote} }  \centering \medskip
	\resizebox{\linewidth}{!}{
 \begin{tabular}{|c c c c c c|}
	\hline
			\multicolumn{6}{|c|}{Estimation of $\alpha_k$} \\
			\hline	
			State & Party & Class & Age &  Gender & Lawyer  \\ 
			\hline
			\textbf{2.342} & 	\textbf{0.167} & 0.021 & 0 & 0 & 0.018 \\
			\textbf{(1.846,2.838)} & 	\textbf{(0.153,0.181)} & (-0.034,0.075) & 0 & 0 & (-0.047,0.083) \\
			\hline
			Executive & Businessman & Farmer &  Army & Teacher & Professor  \\ \hline
			0 & \textbf{0.451} & 0 & 0 & 0 & 0  	 \\ 	
			0 & \textbf{(0.018,0.884)} & 0 & 0 & 0 & 0  \\
			\hline
			Tweets & Followers & Twitter Follower-Followee Relationship &  &  &   \\ \hline
			\textbf{-0.091} & 0 & \textbf{0.144} & & & \\ 
			\textbf{(-0.136,-0.046)} & 0 & \textbf{(0.110,0.177)} & & &  \\
			\hline
		\end{tabular}
    } \vspace{-1em}
\end{table}
	
There is statistically clear evidence of a positive association between the Twitter follower-followee relationship adjacency matrix and senator voting patterns. A possible explanation is that one senator who follows the other senator on Twitter has more exposure to their advocated ideologies, and hence is more likely to vote similarly. It is also possible that the senator who follows the other senator on Twitter already agrees with their ideologies in the first place, while their interactions on Twitter further reinforce such agreement, leading to positive associations in their voting patterns. The similarity in terms of senators' popularity on Twitter, as measured by the number of followers, does not have any effect on the association between senators' votes. Interestingly, although the effect of the senators' number of tweets is found to be negative, when we run a separate analysis based on fitting an Ising similarity regression model with only this similarity matrix, its associated coefficient becomes positive: $\hat{\alpha}_{\textrm{Tweets}} = 0.083$ with 95\% confidence interval being $(0.080,0.087)$.  Therefore, the negative coefficient found in Table \ref{table:rollcall_vote} is conjectured to be due to a large amount of information contained within this similarity matrix that could be explained by other similarity matrices i.e., a form of matrix multicollinearity.

 \begin{figure}[t!]
	\begin{center}
	    \includegraphics[width=0.75\textwidth]{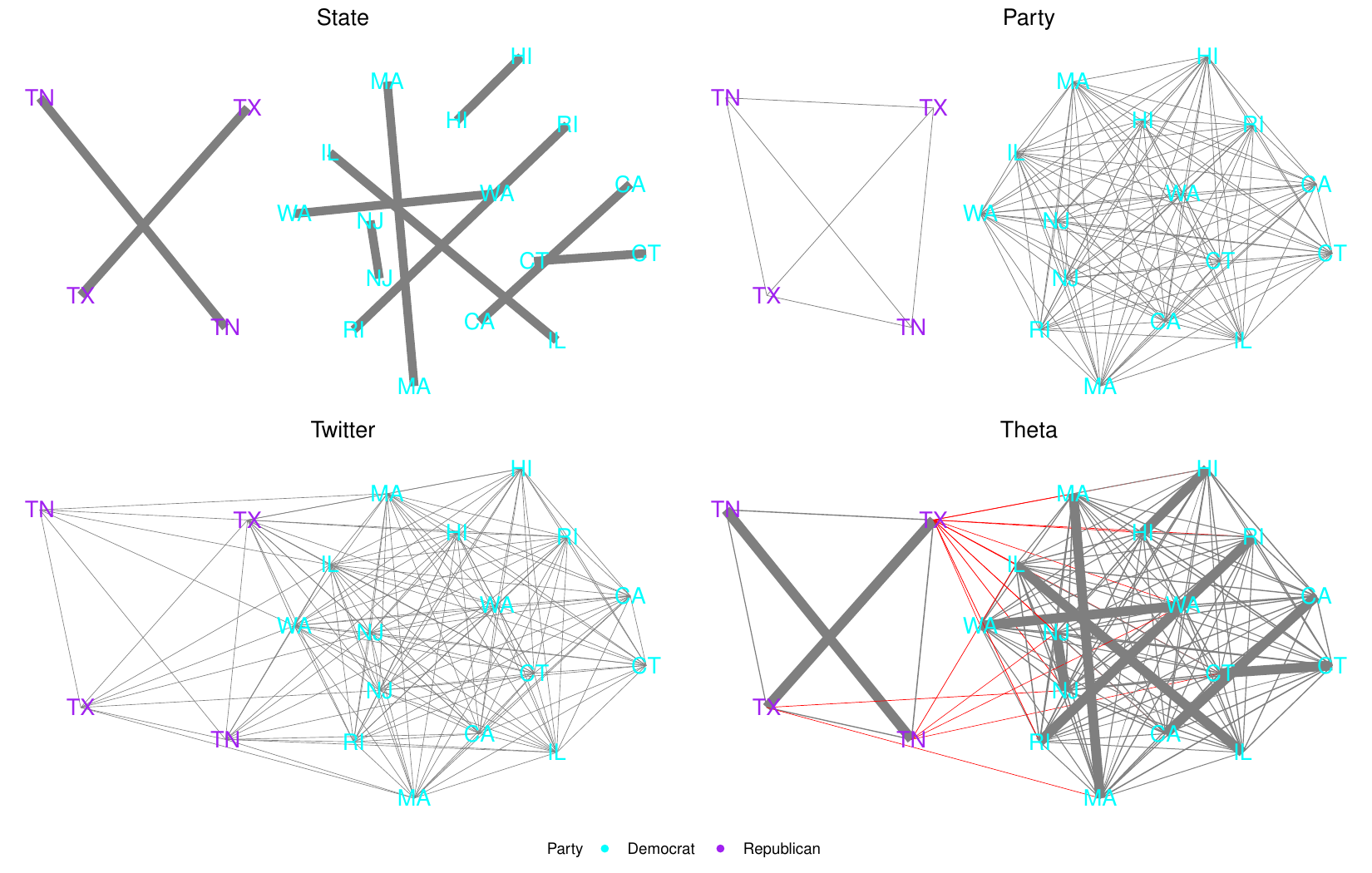}\par
	\end{center}
		\caption{Graphs of weighted similarity/adjacency  matrices $\hat{\alpha}_k \bmW_k$ associated with state, party and Twitter follower-followee relationship for a subset of 20 senators, where the edge width is proportional to the estimated $\hat{\alpha}_k$. The bottom right plot presents the weighted graph based on the estimated interaction matrix $\hat{\bm \Theta}$, where the edge width is proportional to its associated $\hat{\theta}_{jj'}$ and edges between senators with different states and parties are colored red. Nodes are labeled with the state abbreviation of each senator and the color represents senator's party. 
		\label{fig:network_plots_rollcall_median}
		} \vspace{-1em}
\end{figure}
	
Figure \ref{fig:network_plots_rollcall_median} presents graphs of the weighted similarity/adjacency matrices $\hat{\alpha}_k \bmW_k$ for selected attributes, together with the estimated interaction matrix $\hat{\bm \Theta}$ describing the voting associations for a subset of 20 senators, where $\hat{\bm \Theta} = \sum_{j=1}^{100}\hat{\theta}_{jj} \bm \Delta_{jj} + \sum_{k=1}^{15}\hat{\alpha}_k \bmW_k$ based on equation \eqref{eq:matrix_regression}, and $\hat{\alpha}_k$ and $\hat{\theta}_{jj}$ are the regularized pseudo-likelihood estimators. The choice of senators to be included is made by beginning with an empty set, and sequentially adding pairs of senators who have the largest off-diagonal elements $\hat{\theta}_{jj'}$ in $\hat{\bm{\Theta}}$ until the set contains the top 20 senators. To further improve clarity of the graph based on the estimated $\hat{\bm{\Theta}}$, we remove all edges with $\hat{\theta}_{jj'} \leq 0.1297 $ (the median of $\{\hat{\theta}_{jj'}: j \neq j'\}$); this explains the visually denser graph of Twitter follower-followee relationship compared to that of $\hat{\bm \Theta}$ in Figure \ref{fig:network_plots_rollcall_median}. \zy{\label{page:edge_removal}We emphasize that the removal of edges here is purely to improve the visualization and interpretability of the graph, noting our focus is on similarity selection and not edge selection as discussed earlier in Sections \ref{sec:intro} and \ref{subsec:estimation}. The corresponding graph without such removal of edges can be found in Figure \ref{fig:dense_theta} in the supplementary material, and provides qualitatively similar conclusions as above.}
It can be seen that the estimated strong voting associations (i.e., large estimated values of $\theta_{jj'}$) between senators are being driven heavily not only by similarities in senators' state and party affiliations, but also their relationships on Twitter as indicated by the red-colored edges in the graph of $\hat{\bm{\Theta}}$. Indeed, the fitted Ising similarity regression model yields a log pseudo-likelihood of -1125.28, compared to -9441.01 for a null model with only the main effect parameters. This corresponds to a pseudo $R^2$ \citep{cohenETAL2013} of approximately 0.88, indicating the selected model provides a much better fit than the null model through the inclusion of the available $\bm{W}_k$ matrices  to explain the dependence structure.
 
\section{Conclusion} \label{sec:conclusion}
We develop an Ising similarity regression model to study the effect of predictor information, arising in the form of similarity matrices, on conditional dependence relationships among binary responses. A computationally efficient regularized pseudo-likelihood estimator using the adaptive lasso penalty is proposed, which we demonstrate to be estimation and model selection consistent under a general setting where the number of similarity matrices $ K $ and responses $p$ grow with sample size $n$. Simulations demonstrate the strong finite sample point estimation and selection performance of the proposed estimator, \zy{\label{page:alternative_estimator_discussion_conclusion} especially compared with several traditional Ising model estimators in recovering the interaction matrix, as it incorporates additional information from relevant similarity matrices. The cross-validation approach for choosing the tuning parameter is shown to have comparable performance to the BIC in terms of model selection.\label{page:alternative_estimator_discussion_conclusion_end}} Applying the proposed model to the U.S. Senate roll call voting data not only identifies well-documented state and party effects in driving senator voting patterns, but also establishes new insights into the importance of senators' similarities in businessman occupation and social network relationships on their voting dependence. These findings are new and differ from those obtained from the graphical structure analyses using standard Ising models \citep{guoETAL2010,guoETAL2015}. In particular, \zy{while graphical structure analyses aim to reveal dependency structures among responses (senators) through edge selection, our method seeks to recover and quantify how different auxiliary information (senator's attributes) influence this structure \label{page:similarity_selection_conclusion}via similarity selection}. While various methods have been developed for the former problem, our approach is one of the first to have been developed for addressing the latter. 

A logical next step would be to extend the proposed model to incorporate predictor information from both observations and responses, thus forming a sort of ``double Ising similarity regression model''. \zy{\label{page:violation_iid_start}This may involve relaxing the assumption of $\bm{y}_1, \cdots, \bm{y}_n \stackrel{i.i.d.}{\sim} f(\cdot; \bmvartheta)$ to consider correlated data such as spatial and/or temporal data, where similarity matrices could be built up based on knowledge of spatial or temporal distance \citep[e.g.,][]{bonat2016} to account for spatial or temporal dependence between observations. The assumption could also be relaxed by extending the Ising similarity regression model to allow for heterogeneous interaction matrices $\bm{\Theta}^{(i)}$ for different $\bm{y}_i$ vectors, e.g. by considering $\bm{W}_k^{(i)}$ that are heterogeneous across $i=1,\cdots,n$; see for instance, \cite{tao2021} who model heterogeneous covariance matrices of continuous response vectors based on linear combination of heterogeneous similarity matrices. \label{page:violation_iid_end}}
    
\zy{\label{page:future_inference_theory}While this article focuses on the estimation and selection of the regression coefficients $\bmalpha$, future research would involve rigorous theoretical study of the inferential component of our method e.g., by establishing asymptotic normality results for the regularized pseudo-likelihood estimator. \label{page:future_IC}Given the comparable model selection performance of the BIC to our cross-validation method, it would also be useful to extend this investigation to consider other types of information criterion in selecting $\lambda$ \citep*[e.g.,][]{zhangETAL2010,fanANDtang2012}.} On a related note, other penalty functions could be used to replace the adaptive lasso penalty for similarity selection, including the use of group or fusion  penalties when the set of similarity matrices exhibits some sort of hierarchy or ordering (\citeauthor*{hui2017hierarchical}, \citeyear{hui2017hierarchical}; \citeauthor{zhang2023cause}, \citeyear{zhang2023cause}). Finally, while this work focuses on binary responses using the Ising model, similar modeling idea could be extended to quadratic exponential families \citep*{gourierouxETAL1984} to allow for other response types.

\section*{Supplementary Materials}
	
The Supplementary Material contains sample versions of Conditions \ref{cond:eigenvalue_pop} and \ref{cond:irrepresent_pop}, proofs of the theorems, \zy{inference method, additional simulation results}, along with supplementary details of application to the U.S. Senate roll call voting data\zy{, as well as an additional application to the Scotland Carabidae ground beetle data.}
	\par
\section*{Acknowledgements}
Zhi Yang Tho was supported by an Australian Government Research Training Program scholarship. Francis KC Hui was supported by an Australian Research Council Discovery Project DP230101908. Thanks to Alan Welsh for useful discussions.
	

\bibhang=1.7pc
\bibsep=2pt
\fontsize{9}{14pt plus.8pt minus .6pt}\selectfont
\renewcommand\bibname{\large \bf References}
\expandafter\ifx\csname
natexlab\endcsname\relax\def\natexlab#1{#1}\fi
\expandafter\ifx\csname url\endcsname\relax
  \def\url#1{\texttt{#1}}\fi
\expandafter\ifx\csname urlprefix\endcsname\relax\def\urlprefix{URL}\fi

\bibliographystyle{chicago}      
\bibliography{bibliography_Sinica}   

\begin{thebibliography}{}

\bibitem[\protect\citeauthoryear{Akaike}{Akaike}{1998}]{akaike1998information}
Akaike, H. (1998).
\newblock Information theory and an extension of the maximum likelihood
  principle.
\newblock In {\em Selected Papers of Hirotugu Akaike}, pp.\  199--213.
  Springer.

\bibitem[\protect\citeauthoryear{Anderson}{Anderson}{1973}]{anderson1973}
Anderson, T.~W. (1973).
\newblock Asymptotically efficient estimation of covariance matrices with
  linear structure.
\newblock {\em The Annals of Statistics\/}~{\em \bf 1}, 135--141.

\bibitem[\protect\citeauthoryear{Banerjee, El~Ghaoui, and
  d'~Aspremont}{Banerjee et~al.}{2008}]{banerjeeETAL2008}
Banerjee, O., L.~El~Ghaoui, and A.~d'~Aspremont (2008).
\newblock {Model selection through sparse maximum likelihood estimation for
  multivariate Gaussian or binary data}.
\newblock {\em Journal of Machine Learning Research\/}~{\em \bf 9}, 485--516.

\bibitem[\protect\citeauthoryear{Bhattacharya and Mukherjee}{Bhattacharya and
  Mukherjee}{2018}]{bhattacharyaANDmukherjee2017}
Bhattacharya, B.~B. and S.~Mukherjee (2018).
\newblock {Inference in Ising models}.
\newblock {\em Bernoulli\/}~{\em \bf 24}, 493--525.

\bibitem[\protect\citeauthoryear{Bonat and Jørgensen}{Bonat and
  Jørgensen}{2016}]{bonat2016}
Bonat, W.~H. and B.~Jørgensen (2016).
\newblock Multivariate covariance generalized linear models.
\newblock {\em Journal of the Royal Statistical Society: Series C (Applied
  Statistics)\/}~{\em \bf 65}, 649--675.

\bibitem[\protect\citeauthoryear{Cheng, Levina, Wang, and Zhu}{Cheng
  et~al.}{2014}]{chengETAL2014}
Cheng, J., E.~Levina, P.~Wang, and J.~Zhu (2014).
\newblock {A sparse Ising model with covariates}.
\newblock {\em Biometrics\/}~{\em \bf 70}, 943--953.

\bibitem[\protect\citeauthoryear{Cohen, Cohen, West, and Aiken}{Cohen
  et~al.}{2013}]{cohenETAL2013}
Cohen, J., P.~Cohen, S.~West, and L.~Aiken (2013).
\newblock {\em Applied multiple regression/correlation analysis for the
  behavioral sciences}.
\newblock Taylor \& Francis.

\bibitem[\protect\citeauthoryear{Fan and Tang}{Fan and
  Tang}{2013}]{fanANDtang2012}
Fan, Y. and C.~Y. Tang (2013).
\newblock Tuning parameter selection in high dimensional penalized likelihood.
\newblock {\em Journal of the Royal Statistical Society. Series B (Statistical
  Methodology)\/}~{\em \textbf{75}}, 531--552.

\bibitem[\protect\citeauthoryear{Friedman, Hastie, and Tibshirani}{Friedman
  et~al.}{2007}]{friedmanETAL2008}
Friedman, J., T.~Hastie, and R.~Tibshirani (2007).
\newblock {Sparse inverse covariance estimation with the graphical lasso}.
\newblock {\em Biostatistics\/}~{\em \bf 9}, 432--441.

\bibitem[\protect\citeauthoryear{Friedman, Hastie, and Tibshirani}{Friedman
  et~al.}{2010}]{glmnetpackage}
Friedman, J., T.~Hastie, and R.~Tibshirani (2010).
\newblock Regularization paths for generalized linear models via coordinate
  descent.
\newblock {\em Journal of Statistical Software\/}~{\em \bf 33}, 1--22.

\bibitem[\protect\citeauthoryear{Gourieroux, Monfort, and Trognon}{Gourieroux
  et~al.}{1984}]{gourierouxETAL1984}
Gourieroux, C., A.~Monfort, and A.~Trognon (1984).
\newblock Pseudo maximum likelihood methods: Theory.
\newblock {\em Econometrica\/}~{\em \bf 52}, 681--700.

\bibitem[\protect\citeauthoryear{Guo, Cheng, Levina, Michailidis, and Zhu}{Guo
  et~al.}{2015}]{guoETAL2015}
Guo, J., J.~Cheng, E.~Levina, G.~Michailidis, and J.~Zhu (2015).
\newblock Estimating heterogeneous graphical models for discrete data with an
  application to roll call voting.
\newblock {\em The Annals of Applied Statistics\/}~{\em \bf 9}, 821--848.

\bibitem[\protect\citeauthoryear{Guo, Levina, Michailidis, and Zhu}{Guo
  et~al.}{2010}]{guoETAL2010}
Guo, J., E.~Levina, G.~Michailidis, and J.~Zhu (2010).
\newblock {Joint structure estimation for categorical Markov networks}.

\bibitem[\protect\citeauthoryear{Hammersley and Clifford}{Hammersley and
  Clifford}{1971}]{hammersleyANDclifford1971}
Hammersley, J.~M. and P.~Clifford (1971).
\newblock Markov fields on finite graphs and lattices.

\bibitem[\protect\citeauthoryear{Hastie, Tibshirani, and Wainwright}{Hastie
  et~al.}{2015}]{hastieETAL2015}
Hastie, T., R.~Tibshirani, and M.~Wainwright (2015).
\newblock {\em Statistical learning with sparsity: The lasso and
  generalizations}.
\newblock Chapman \& Hall/CRC Monographs on Statistics \& Applied Probability.
  CRC Press.

\bibitem[\protect\citeauthoryear{H{{\"o}}fling and Tibshirani}{H{{\"o}}fling
  and Tibshirani}{2009}]{hoflingANDtibshirani2009}
H{{\"o}}fling, H. and R.~Tibshirani (2009).
\newblock {Estimation of sparse binary pairwise Markov networks using
  pseudo-likelihoods}.
\newblock {\em Journal of Machine Learning Research\/}~{\em \textbf{10}},
  883--906.

\bibitem[\protect\citeauthoryear{Huang, Ma, and Zhang}{Huang
  et~al.}{2008}]{huang2008adaptive}
Huang, J., S.~Ma, and C.-H. Zhang (2008).
\newblock {Adaptive lasso for sparse high-dimensional regression models}.
\newblock {\em Statistica Sinica\/}~{\em \bf 18}, 1603--1618.

\bibitem[\protect\citeauthoryear{Hui, M{\"u}ller, and Welsh}{Hui
  et~al.}{2017a}]{hui2017hierarchical}
Hui, F. K.~C., S.~M{\"u}ller, and A.~H. Welsh (2017a).
\newblock {Hierarchical selection of fixed and random effects in generalized
  linear mixed models}.
\newblock {\em Statistica Sinica\/}~{\em \bf 27}, 501--518.

\bibitem[\protect\citeauthoryear{Hui, M{\"u}ller, and Welsh}{Hui
  et~al.}{2017b}]{hui2017joint}
Hui, F. K.~C., S.~M{\"u}ller, and A.~H. Welsh (2017b).
\newblock {Joint selection in mixed models using regularized PQL}.
\newblock {\em Journal of the American Statistical Association\/}~{\em \bf
  112}, 1323--1333.

\bibitem[\protect\citeauthoryear{Hui, M{\"u}ller, and Welsh}{Hui
  et~al.}{2018}]{hui2018sparse}
Hui, F. K.~C., S.~M{\"u}ller, and A.~H. Welsh (2018).
\newblock {Sparse pairwise likelihood estimation for multivariate longitudinal
  mixed models}.
\newblock {\em Journal of the American Statistical Association\/}~{\em \bf
  113}, 1759--1769.

\bibitem[\protect\citeauthoryear{Hui, M{\"u}ller, and Welsh}{Hui
  et~al.}{2023}]{huiETAL2023gee}
Hui, F. K.~C., S.~M{\"u}ller, and A.~H. Welsh (2023).
\newblock {GEE-assisted variable selection for latent variable models with
  multivariate binary data}.
\newblock {\em Journal of the American Statistical Association\/}~{\em \bf
  118}, 1252--1263.

\bibitem[\protect\citeauthoryear{Ising}{Ising}{1925}]{ising1925}
Ising, E. (1925).
\newblock Beitrag zur theorie der ferromagnetismus.
\newblock {\em Zeitschrift für Physik\/}~{\em \bf 31}, 253--258.

\bibitem[\protect\citeauthoryear{Johnson and Wichern}{Johnson and
  Wichern}{1992}]{johnsonANDwichern1992}
Johnson, R.~A. and D.~W. Wichern (1992).
\newblock {\em Applied multivariate statistical analysis}.
\newblock Prentice Hall.

\bibitem[\protect\citeauthoryear{Lee and Xue}{Lee and
  Xue}{2018}]{leeANDxue2018}
Lee, K.~H. and L.~Xue (2018).
\newblock {Nonparametric finite mixture of Gaussian graphical models}.
\newblock {\em Technometrics\/}~{\em \bf 60}, 511--521.

\bibitem[\protect\citeauthoryear{Lee, Ganapathi, and Koller}{Lee
  et~al.}{2006}]{leeETAL2006}
Lee, S.-I., V.~Ganapathi, and D.~Koller (2006).
\newblock {Efficient structure learning of Markov networks using $L_1$-
  regularization}.
\newblock In B.~Sch\"{o}lkopf, J.~Platt, and T.~Hoffman (Eds.), {\em Advances
  in Neural Information Processing Systems}, Volume \textbf{19}. MIT Press.

\bibitem[\protect\citeauthoryear{Liu, Chen, Wasserman, and Lafferty}{Liu
  et~al.}{2010}]{liuETAL2010}
Liu, H., X.~Chen, L.~Wasserman, and J.~Lafferty (2010).
\newblock Graph-valued regression.
\newblock In J.~Lafferty, C.~Williams, J.~Shawe-Taylor, R.~Zemel, and
  A.~Culotta (Eds.), {\em Advances in Neural Information Processing Systems},
  Volume \textbf{23}. Curran Associates, Inc.

\bibitem[\protect\citeauthoryear{Majewski, Li, and Ott}{Majewski
  et~al.}{2001}]{majewskiETAL2001}
Majewski, J., H.~Li, and J.~Ott (2001).
\newblock {The Ising model in physics and statistical genetics}.
\newblock {\em The American Journal of Human Genetics\/}~{\em \bf 69},
  853--862.

\bibitem[\protect\citeauthoryear{McElroy and Trimbur}{McElroy and
  Trimbur}{2023}]{mcelroyANDtrimbur2023}
McElroy, T.~S. and T.~Trimbur (2023).
\newblock Variable targeting and reduction in large vector autoregressions with
  applications to workforce indicators.
\newblock {\em Journal of Applied Statistics\/}~{\em \bf 50}, 1515--1537.

\bibitem[\protect\citeauthoryear{Meinshausen and Bühlmann}{Meinshausen and
  Bühlmann}{2006}]{meinhausenANDbuhlmann2006}
Meinshausen, N. and P.~Bühlmann (2006).
\newblock High-dimensional graphs and variable selection with the lasso.
\newblock {\em The Annals of Statistics\/}~{\em \bf 34}, 1436--1462.

\bibitem[\protect\citeauthoryear{Nghiem, Hui, M{\"u}ller, and Welsh}{Nghiem
  et~al.}{2022}]{nghiem2022sparse}
Nghiem, L.~H., F.~K.~C. Hui, S.~M{\"u}ller, and A.~H. Welsh (2022).
\newblock {Sparse sliced inverse regression via Cholesky matrix penalization}.
\newblock {\em Statistica Sinica\/}~{\em \bf 32}, 2431--2453.

\bibitem[\protect\citeauthoryear{Ni, Stingo, and Baladandayuthapani}{Ni
  et~al.}{2022}]{niETAL2022}
Ni, Y., F.~C. Stingo, and V.~Baladandayuthapani (2022).
\newblock {Bayesian covariate-dependent Gaussian graphical models with varying
  structure}.
\newblock {\em Journal of Machine Learning Research\/}~{\em \bf 23}, 1--29.

\bibitem[\protect\citeauthoryear{Parlett}{Parlett}{1980}]{parlett1980}
Parlett, B. (1980).
\newblock {\em The symmetric eigenvalue problem}.
\newblock Prentice-Hall International Series in Computer Science.
  Prentice-Hall.

\bibitem[\protect\citeauthoryear{Pourahmadi}{Pourahmadi}{1999}]{pourahmadi1999}
Pourahmadi, M. (1999).
\newblock Joint mean-covariance models with applications to longitudinal data:
  Unconstrained parameterisation.
\newblock {\em Biometrika\/}~{\em \bf 86}, 677--690.

\bibitem[\protect\citeauthoryear{Ravikumar, Wainwright, and Lafferty}{Ravikumar
  et~al.}{2010}]{ravikumarETAL2010}
Ravikumar, P., M.~J. Wainwright, and J.~D. Lafferty (2010).
\newblock {High-dimensional Ising model selection using $l_1$-regularized
  logistic regression}.
\newblock {\em The Annals of Statistics\/}~{\em \bf 38}, 1287--1319.

\bibitem[\protect\citeauthoryear{Schwarz}{Schwarz}{1978}]{schwarz1978}
Schwarz, G. (1978).
\newblock {Estimating the dimension of a model}.
\newblock {\em The Annals of Statistics\/}~{\em \bf 6}, 461--464.

\bibitem[\protect\citeauthoryear{Tsay}{Tsay}{2013}]{tsay2013multivariate}
Tsay, R. (2013).
\newblock {\em Multivariate time series analysis: With R and financial
  applications}.
\newblock Wiley Series in Probability and Statistics. Wiley.

\bibitem[\protect\citeauthoryear{Wainwright}{Wainwright}{2019}]{wainwright2019}
Wainwright, M. (2019).
\newblock {\em {High-dimensional statistics: A non-asymptotic viewpoint}}.
\newblock Cambridge Series in Statistical and Probabilistic Mathematics.
  Cambridge University Press.

\bibitem[\protect\citeauthoryear{Wang, Li, and Leng}{Wang
  et~al.}{2009}]{wang2009shrinkage}
Wang, H., B.~Li, and C.~Leng (2009).
\newblock {Shrinkage tuning parameter selection with a diverging number of
  parameters}.
\newblock {\em Journal of the Royal Statistical Society: Series B (Statistical
  Methodology)\/}~{\em \bf 71}, 671--683.

\bibitem[\protect\citeauthoryear{Wang, Baladandayuthapani, Kaseb, Amin, Hassan,
  Wang, and Morris}{Wang et~al.}{2022}]{wangETAL2022}
Wang, Z., V.~Baladandayuthapani, A.~O. Kaseb, H.~M. Amin, M.~M. Hassan,
  W.~Wang, and J.~S. Morris (2022).
\newblock Bayesian edge regression in undirected graphical models to
  characterize interpatient heterogeneity in cancer.
\newblock {\em Journal of the American Statistical Association\/}~{\em \bf
  117}, 533--546.

\bibitem[\protect\citeauthoryear{Warton, Thibaut, and Wang}{Warton
  et~al.}{2017}]{warton2017pit}
Warton, D.~I., L.~Thibaut, and Y.~A. Wang (2017).
\newblock {The PIT-trap -- A “model-free” bootstrap procedure for inference
  about regression models with discrete, multivariate responses}.
\newblock {\em PloS one\/}~{\em \bf 12}, e0181790.

\bibitem[\protect\citeauthoryear{Whittaker}{Whittaker}{1990}]{whittaker1990}
Whittaker, J. (1990).
\newblock {\em Graphical models in applied multivariate statistics}.
\newblock Wiley Series in Probability and Statistics. Wiley.

\bibitem[\protect\citeauthoryear{Xue, Zou, and Cai}{Xue
  et~al.}{2012}]{xueETAL2012}
Xue, L., H.~Zou, and T.~Cai (2012).
\newblock {Nonconcave penalized composite conditional likelihood estimator of
  sparse Ising models}.
\newblock {\em The Annals of Statistics\/}~{\em \bf 40}, 1403--1429.

\bibitem[\protect\citeauthoryear{Yuan and Lin}{Yuan and
  Lin}{2007}]{yuanANDlin2007}
Yuan, M. and Y.~Lin (2007).
\newblock Model selection and estimation in the gaussian graphical model.
\newblock {\em Biometrika\/}~{\em \bf 94}, 19--35.

\bibitem[\protect\citeauthoryear{Zhang, Huang, Hui, and Haberman}{Zhang
  et~al.}{2023}]{zhang2023cause}
Zhang, X., F.~Huang, F.~K.~C. Hui, and S.~Haberman (2023).
\newblock {Cause-of-death mortality forecasting using adaptive penalized tensor
  decompositions}.
\newblock {\em Insurance: Mathematics and Economics\/}~{\em \bf 111}, 193--213.

\bibitem[\protect\citeauthoryear{Zhang, Li, and Tsai}{Zhang
  et~al.}{2010}]{zhangETAL2010}
Zhang, Y., R.~Li, and C.-L. Tsai (2010).
\newblock Regularization parameter selections via generalized information
  criterion.
\newblock {\em Journal of the American Statistical Association\/}~{\em
  \textbf{105}}, 312--323.

\bibitem[\protect\citeauthoryear{Zou}{Zou}{2006}]{zou2006}
Zou, H. (2006).
\newblock The adaptive lasso and its oracle properties.
\newblock {\em Journal of the American Statistical Association\/}~{\em \bf
  101}, 1418--1429.

\bibitem[\protect\citeauthoryear{Zou, Lan, Li, and Tsai}{Zou
  et~al.}{2022}]{tao2021}
Zou, T., W.~Lan, R.~Li, and C.-L. Tsai (2022).
\newblock Inference on covariance-mean regression.
\newblock {\em Journal of Econometrics\/}~{\em \bf 230}, 318--338.

\bibitem[\protect\citeauthoryear{Zou, Lan, Wang, and Tsai}{Zou
  et~al.}{2017}]{tao2017}
Zou, T., W.~Lan, H.~Wang, and C.-L. Tsai (2017).
\newblock Covariance regression analysis.
\newblock {\em Journal of the American Statistical Association\/}~{\em \bf
  112}, 266--281.

\bibitem[\protect\citeauthoryear{Zou, Luo, Lan, and Tsai}{Zou
  et~al.}{2020}]{tao2020}
Zou, T., R.~Luo, W.~Lan, and C.-L. Tsai (2020).
\newblock Covariance regression model for non-normal data.
\newblock In C.~F. Lee and J.~C. Lee (Eds.), {\em Handbook of Financial
  Econometrics, Mathematics, Statistics, and Machine Learning}, Chapter 113,
  pp.\  3933--3945. World Scientific.

\end{thebibliography}

\vskip .63cm
\noindent
Zhi Yang Tho
\vskip 2pt
\noindent
The Australian National University, Canberra, ACT 2600, Australia.
\vskip 2pt
\noindent
E-mail: zhiyang.tho@anu.edu.au
\vskip 2pt

\noindent
Francis K.C. Hui
\vskip 2pt
\noindent
The Australian National University, Canberra, ACT 2600, Australia.
\vskip 2pt
\noindent
E-mail: francis.hui@anu.edu.au
\vskip 2pt

\noindent
Tao Zou
\vskip 2pt
\noindent
The Australian National University, Canberra, ACT 2600, Australia.
\vskip 2pt
\noindent
E-mail: tao.zou@anu.edu.au

\end{document}